\documentclass[12pt]{article}
\usepackage{amsfonts,amsmath,amssymb}
\usepackage{color}
\def\C{\mathbb{C}}

\def\bq{ \begin{equation} }
\def\eq{ \end{equation} }
\def\ben{ \begin{eqnarray} }
\def\en{ \end{eqnarray} }

\def\frac#1#2{{#1\over #2}}

\def\on#1#2{\mathop{\vbox{\ialign{##\crcr\noalign{\kern2pt}
$\scriptstyle{#2}$\crcr\noalign{\kern2pt\nointerlineskip}
\kern-2pt$\hfil\displaystyle{#1}\hfil$\crcr}}}\limits}

\def\ldb{\mathopen{\{\!\!\{}} \def\rdb{\mathclose{\}\!\!\}}}

\begin{document}

%%%%%%%%%%%%%%%%%%%%%%%%%%%%%%%%%%%%%%
\baselineskip=15pt
%\begin{flushright}
%Draft\\
%12/04/2003
%\end{flushright}
\vspace{1cm} \centerline{\LARGE  \textbf{Double Poisson
brackets }} 
\vskip0.4cm 
\centerline{\LARGE  \textbf{  on free associative algebras}}

\vskip1cm \hfill
\begin{minipage}{13.5cm}
\baselineskip=15pt
{\bf A Odesskii  ${}^{1}$,
    V Rubtsov ${}^{2,}$ and
   V Sokolov ${}^{3}$} \\ [2ex]
{\footnotesize
${}^1$  Brock University (Canada)
\\
${}^2$ Institute for Theoretical and Experimental Physics, Moscow (Russia) and LAREMA, UMR 6093 du CNRS, Angers University (France)
\\
${}^{3}$ Landau Institute for Theoretical Physics, Moscow, Russia }\\

\vskip2cm{\em to the memory of S.V. Manakov}

\vskip1cm{\bf Abstract}
%\baselineskip=3D15pt

We discuss double Poisson structures in sense of M. Van den Bergh on free associative algebras focusing on
the case of quadratic Poisson brackets. We establish their relations with an associative version of  Young-Baxter equations, we study a bi-hamiltonian property of the linear-quadratic pencil of the double Poisson structure  and propose a classification of the quadratic double Poisson brackets in the case of the algebra with two free generators. Many new examples of quadratic double Poisson brackets are proposed.

\end{minipage}

\vglue1cm \textbf{Address}:
Landau Institute for Theoretical Physics, Kosygina 2, 119334, Moscow, Russia

\textbf{E-mail}:
sokolov@itp.ac.ru, \, odesskii@itp.ac.ru, \, volodya@univ-angers.fr \newpage

\section{Introduction}

 A Poisson structure on a commutative algebra $A$ is a Lie algebra structure on $A$ given
by a Lie bracket
$$
\{ -,- \}:A\times A \mapsto A,
$$
which is a derivation of $A$
i.e. satisfies the Leibniz rule
$$\{a , bc\} = \{a , b\}c+ b\{a , c\},\quad a,b,c\in A,$$
for the right (and, hence, for the left) argument.

It is well-known (see the discussion in \cite{CB}) that a naive translation of this definition
to the case of a non-commutative associative algebra $A$ is not very interesting because of lack
of examples different from the usual commutator (for prime rings it was shown in \cite{FL}).

It turns out \cite{CB,VdB} that a natural generalization of Poisson structures on comutative associative algebras to a non-commutative case  is a Lie structure on the vector space 
$H_0(A,A) = A/[A,A],$ where $[A,A]$ is the vector space spaned by all comutators $a b-b a$ where $a,b\in A.$ The elements of this space are $0-$dimensional cyclic homology classes of $A$ and 
they are represented by "cyclic words" whose letters are the elements of $A$ \cite{Ginzb}. In \cite{CB} such a structure was called an $H_0-$Poisson structure while in \cite{OdRubSok1} 
the terminology  a "non-abelian Poisson bracket" was used. Both names are somehow misleading to our mind (because one deals with a Lie structure with no multiplication structure on $H_0(A,A)$). 
Therefore we would suggest to call it  a {\it trace bracket} by the following reason.

Let $A$ be a (unital) associative algebra over $\C$.  For a fixed natural $n$ we denote by 
$${\rm Rep}_n (A) := {\rm Hom} (A,{\rm Mat_n(\C)})$$ 
the space of $n-$dimensional representations of $A$, and by  $\C[{\rm Rep}_n (A)]$ the coordinate ring of this affine scheme ${\rm Rep}_n (A).$ Let ${\rm tr}: A\to \C[{\rm Rep}_n (A)]$ be the 
trace map. It is clear that ${\rm tr}(a)$ is  a $GL_n(\C)-$invariant element for any $a\in A.$  

Since the map ${\rm tr}$ is well defined also on elements of $A/[A,A],$  any trace bracket $\{ \}$ induces the following genuine Poisson bracket on the representation space function algebras:
$$
\{{\rm tr}(a),{\rm tr}(b)\}={\rm tr}(\{a,b\})
$$
on ${\rm Im} ({\rm tr}).$  This bracket (according to results from \cite{CProc,Rzm} can be extended to the subalgebra of all $GL_n(\C)-$invariant elements of  $\C[{\rm Rep}_n (A)]$ and, 
in certain important cases, even to the whole algebra  $\C[{\rm Rep}_n (A)].$ We shall refer such brackets on  $\C[{\rm Rep}_n (A)]$ and  $\C[{\rm Rep}_n (A)]^{GL_n(\C)}$ as \emph{trace 
Poisson brackets.}

In this paper we shall consider as a basic example the case of free associative algebra
$A=\C<x_1,\ldots,x_m>.$ 
 
The coordinate algebra $\C[{\rm Rep}_n (A)]$ in this case is the polynomial ring of $mn^2$
variables $x^j_{i,\alpha},$ where
$$x_{\alpha} \to M_{\alpha}=\left( \begin{array}{ccc}
x^1_{1,\alpha}& \cdot& x^n_{1,\alpha}\\
\cdot& \cdot& \cdot\\
x^1_{n,\alpha}& \cdot& x^n_{n,\alpha}
\end{array}\right), \qquad  1\leq \alpha\leq m. $$

The map ${\rm tr}$ gives the following interpretation of the variables  $x^j_{i,\alpha}: $ if $E^j_{i}$ denotes
the $(i,j)-$matrix unit (i.e. the $n\times n$ matrix with 0 everywhere except the $i-$th row and $j-$th
column)  then $x^j_{i,\alpha} = {\rm tr}(E^i_{j}M_{\alpha}).$ 

The group $GL_n(\C)$ acts on $M_1,...,M_m$ by the conjugations.
 Any trace  bracket on the free algebra $A$ is extended on $\C[{\rm Rep}_n (A)]$ and it yields a usual $GL_n(\C)-$ invariant Poisson bracket such that the bracket between traces of any 
two matrix polynomials $P_i(M_1,...,M_m), \, i=1,2$ is a trace of some matrix 
polynomial $P_3$.   Notice that not any  $GL_n(\C)-$invariant Poisson bracket on $\C[{\rm Rep}_n (A)]$  is a trace Poisson.

 There are two different ways to represent explicitely the same trace brackets in the free algebra case. One is  a standard way used in the Integrable System theory (see \cite{MikSok}), where  
brackets are given by 
$$  \{a,b\}=<\mbox{grad}\, a, \,  \Theta ( \mbox{grad}\, b)>  ,\ \ \ \
a,b\in A/[A,A], $$
for some Hamiltonian operator $\Theta$, a skew-symmetric operator expressed via left and right multiplication operators on $A.$ The trace brackets define a Hamiltonian formalism for 
integrable models with matrix variables \cite{MikSok}. In particular, some of such models are bi-Hamiltonian with respect to compatible linear and quadratic trace Poisson brackets \cite{OdRubSok1}. 

Another approach can be developped in terms of \emph{double Poisson brackets} introduced in \cite{VdB}. We shall remind their definition.

{\bf Definition}  (M. Van den Bergh). A double Poisson bracket on an associative algebra $A$ is a $\C$-linear map
$\ldb,\rdb : A \otimes A \mapsto  A \otimes A$ satisfying the following conditions:
\begin{equation} \label{dub1}
\ldb u, v\rdb = - \ldb v,u\rdb ^{\circ},
\end{equation}
\begin{equation} \label{dub2}
 \ldb u, \ldb v,w \rdb \rdb_l + \sigma  \ldb v, \ldb w,u \rdb \rdb_l +\sigma^2  \ldb w, \ldb u,v \rdb \rdb_l  =0,
\end{equation}
and
\begin{equation}\label{dub3}
\ldb u, vw \rdb = (v\otimes 1)  \ldb u,w \rdb  + \ldb u,v \rdb (1\otimes  w).
\end{equation}
Here $(u\otimes v)^{\circ}:=v\otimes u$; $\ldb v_1, v_2\otimes v_3 \rdb_l :=\ldb v_1, v_2 \rdb \otimes v_3$
 and $\sigma( v_1 \otimes v_2 \otimes v_3 ):= v_{3}\otimes v_{1} \otimes v_{2}$.
 
 Notice that very similar relations but with a different bi-module structure  in (\ref{dub3})  have appeared in \cite{razum}. 
 
 The relations between double and trace Poisson brackets are established by M.Van den Bergh  \cite{VdB} as follows. 
Let $\mu$ denote the multiplication map $\mu:  A\otimes A \to A$ i.e. $\mu(u\otimes v)=u v.$
We define a $\C-$bilinear bracket operation in $A$ by  $\{-,-\}:=\mu(\ldb -, - \rdb).$

{\bf Proposition 1.} Let $\ldb -, - \rdb$ be a double Poisson bracket on $A$. Then $\{-,-\}$ is a trace bracket on $A/[A, A]$ which is defined as
\begin{equation}\label{trbr}
\lbrace \bar a, \bar b\rbrace =\overline{\mu(\ldb a,b\rdb},
\end{equation}
ou $\bar a$ means the image of $a\in A$ under the natural projection $A\to A/[A,A].$

If $A=\C<x_1,\ldots,x_m>$  is the free associative algebra, then $\C[{\rm Rep}_n(A)] = \C[x_{i,\alpha}^j]$
where $1\leq \alpha \leq m.$ 

If $\ldb x_{\alpha},x_{\beta}\rdb$ is a double Poisson bracket on $A=\C<x_1,\ldots,x_m>$, then, using the Sweedler 
convention and drop the sign of sum, we obtain the trace Poisson brackets on $\C[{\rm Rep}_n(A)] $:
$$\lbrace x_{i,\alpha}^j,x_{k,\beta}^l\rbrace = \ldb x_{\alpha},x_{\beta}\rdb_{k}^{'j}\ldb x_{\alpha},x_{\beta}\rdb_{i}^{"l}$$

In this paper we shall  consider linear and quadratic double Poisson brackets on free associative algebras. It turns out that linear double brackets are in one-to-one correspondence 
with $m$-dimensional associative algebra structures \cite{PVdW}.  We establish relations between a class of quadratic double brackets and constant solutions of classical associative 
Yang-Baxter equation on $Mat_m(\C)$ introduced in \cite{Agui}. The examples of double brackets related to non-constant solutions of various associative Yang-Baxter equations will 
be discussed in the forthcoming paper \cite{OdRubSok2}.

 \section{Quadratic double Poisson brackets}

Let $A=\C<x_1,\ldots,x_m>$ be the free associative algebra. If double brackets $\ldb x_i, x_j \rdb$ between all generators are fixed, then the bracket between two arbitrary elements of $A$ 
is uniquely defined by identities (\ref{dub1}) and (\ref{dub3}). It follows from  (\ref{dub1}) that constant, linear, and quadratic double brackets are defined by 
\begin{equation} \label{dconst}
\ldb x_i,x_j\rdb = c_{ij} 1\otimes 1, \qquad c_{i,j}=-c_{j,i},
 \end{equation}
\begin{equation} \label{dlin}
\ldb x_i,x_j\rdb = b_{ij}^k x_k\otimes 1 - b_{ji}^k1\otimes x_k,
 \end{equation}
and  
\begin{equation} \label{dquad}
\ldb x_{\alpha}, x_{\beta}\rdb =r_{\alpha \beta}^{u v} \, x_u \otimes x_v+a_{\alpha \beta}^{v u} \, x_u x_v\otimes 1-a_{\beta \alpha}^{u v} \,1\otimes  x_v x_u,   \end{equation}
where
\begin{equation}\label{r1}
r^{\sigma \epsilon}_{\alpha\beta}=-r^{\epsilon\sigma}_{\beta\alpha},
\end{equation}
correspondingly.  The summation with respect to repeated indexes is assumed.

It is easy to verify that the bracket (\ref{dconst}) satisfies (\ref{dub2}) for any skew-symmetric tensor $c_{ij}$. For the bracket (\ref{dlin}) the condition (\ref{dub2}) is equivalent 
to the identity 
\begin{equation}\label{r0}
b^{\mu}_{\alpha \beta} b^{\sigma}_{\mu \gamma}=b^{\sigma}_{\alpha \mu} b^{\mu}_{\beta \gamma},
\end{equation}
which means that  $b^{\sigma}_{\alpha \beta}$
are structure constants of an associative algebra ${\cal A}$.  

{\bf Proposition 2.} The bracket (\ref{dquad}) satisfies (\ref{dub2}) iff the following relations hold:

\begin{equation}\label{r2}
r^{\lambda\sigma}_{\alpha\beta}
r^{\mu\nu}_{\sigma\tau}+r^{\mu\sigma}_{\beta\tau} r^{\nu\lambda}_{\sigma\alpha}+r^{\nu\sigma}_{\tau\alpha} r^{\lambda\mu}_{\sigma\beta}=0,
\end{equation}

\begin{equation}\label{r3}
a^{\sigma\lambda}_{\alpha\beta} a^{\mu\nu}_{\tau\sigma}=a^{\mu\sigma}_{\tau\alpha} a^{\nu\lambda}_{\sigma\beta},
\end{equation}
\begin{equation}\label{r4}
a^{\sigma\lambda}_{\alpha\beta} a^{\mu\nu}_{\sigma\tau}=a^{\mu\sigma}_{\alpha\beta} r^{\lambda\nu}_{\tau\sigma}+a^{\mu\nu}_{\alpha\sigma}
r^{\sigma\lambda}_{\beta\tau}
\end{equation}
and
\begin{equation}\label{r5}
a^{\lambda\sigma}_{\alpha\beta} a^{\mu\nu}_{\tau\sigma}=a^{\sigma\nu}_{\alpha\beta} r^{\lambda\mu}_{\sigma\tau}+a^{\mu\nu}_{\sigma\beta}
r^{\sigma\lambda}_{\tau\alpha}.
\end{equation}

The trace Poisson bracket corresponding to any double Poisson bracket (\ref{dquad})  can be defined on  $\C[{\rm Rep}_n (A)]$ by the following way \cite{OdRubSok1}: 
\begin{equation}\label{Poisson}
\{x^j_{i,\alpha},x^{j^{\prime}}_{i^{\prime},\beta}\}=
r^{\gamma\epsilon}_{\alpha\beta}x^{j^{\prime}}_{i,\gamma}x^j_{i^{\prime},\epsilon}+
a^{\gamma\epsilon}_{\alpha\beta}x^k_{i,\gamma}x^{j^{\prime}}_{k,\epsilon}\delta^j_{i^{\prime}}-
a^{\gamma\epsilon}_{\beta\alpha}x^k_{i^{\prime},\gamma}x^{j}_{k,\epsilon}\delta^{j^{\prime}}_i
\end{equation}
where $x^j_{i,\alpha}$ are entries of the matrix $x_{\alpha}$ and $\delta^{j}_i$ is the Kronecker delta-symbol. Relations  (\ref{r1}), (\ref{r2})-(\ref{r5}) hold iff (\ref{Poisson}) 
is a Poisson bracket.

Under a linear change of the generators $x_\alpha \to g^\beta_\alpha x_\beta$ the coefficients of tensors $r$ and $a$ are
transformed in the standard way:
\begin{equation}\label{basis}
r_{\alpha\beta}^{\gamma \sigma} \to g^{\lambda}_\alpha g^{\mu}_\beta h_{\nu}^\gamma h_{\epsilon}^\sigma
\,r_{\lambda \mu}^{\nu \epsilon}, \qquad
a_{\alpha\beta}^{\gamma\sigma} \to g^{\lambda}_\alpha g^{\mu}_\beta h_{\nu}^\gamma h_{\epsilon}^\sigma
\,a_{\lambda \mu}^{\nu \epsilon},
\end{equation}
Here $g_\alpha^\beta h_\beta^\gamma=\delta_\alpha^\gamma$.

The system  of algebraic equations (\ref{r1}), (\ref{r2})-(\ref{r5})  admits the following
involution:
\begin{equation}\label{tran3} r^{\gamma\sigma}_{\alpha\beta}\to r^{\sigma \gamma}_{\alpha\beta},\qquad
a^{\gamma\sigma}_{\alpha\beta}\to -a^{\sigma\gamma}_{\beta\alpha}. \end{equation}

Given a solution $r$ of (\ref{r1}), (\ref{r2}), one can put $a^{ij}_{uv}=0$ to satisfy equations (\ref{r3})-(\ref{r5}). Note that the algebraic system of equations (\ref{r1}), 
(\ref{r2}) besides (\ref{tran3}) admits the involution 
\begin{equation}\label{tran1} r^{\gamma\sigma}_{\alpha\beta}\to r_{\gamma\sigma}^{\alpha\beta}.
\end{equation}

Some examples of double Poisson brackets with zero tensor $a$ can be found using the one-to-one correspondence \cite{Agui} between  
solutions of (\ref{r1}), (\ref{r2}) up to equivalence (\ref{basis}) and exact
representations of anti-Frobenius algebras up to isomorphisms. 

Recall that an {\it anti-Frobenius algebra} is an associative algebra  $\cal J$
(not necessarily with unity) with non-degenerate
anti-symmetric bilinear form $(~,~)$ satisfying the following relation
\begin{equation}\label{af}
(x,yz)+(y,zx)+(z,xy)=0
\end{equation}
for all $x,y,z\in \cal J$.  

{\bf Construction.}   Let
 $\cal J$ be a $p$-dimensional 
associative anti-Frobenius subalgebra in $Mat_m$ with a basis $y_i=(y^{\alpha}_{\gamma,i}), \, i=1,...,p$. 
Let $G=(g_{ij})$ be the matrix of the form.  Then the tensor
$r^{\alpha \beta}_{\gamma \delta}=g^{i j}y^{\alpha}_{\gamma,i}y^{\beta}_{\delta,j},$
where $G^{-1}=(g^{ij})$, satisfies (\ref{r1}), (\ref{r2}).  
 
 {\bf Example 1}. Let $\cal J$ be the associative algebra of all $m\times m$ matrices with zero $m$-th row,  $(x,y)={\rm trace}([x,y]\, k^T),$ where $k\in {\cal J}$ is a fixed generic 
element. The corresponding bracket up to equivalence (\ref{basis}) is given by a tensor $r$ with the following  non-zero components 
\begin{equation}\label{ex4}
  r^{\alpha \beta}_{\alpha \beta}=r^{\beta \alpha}_{\alpha \beta}=r^{\alpha\alpha}_{\beta\alpha}=-r^{\alpha\alpha}_{\alpha\beta}=\frac{1}{\lambda_\alpha-\lambda_\beta},
\quad \alpha\ne \beta.
\end{equation}
Here $\lambda_1,\ldots,\lambda_N$ are arbitrary pairwise distinct parameters. The generalization of (\ref{ex4}) to the case $k$ zero rows, where $k$ is any divisor of $m$, 
can be found in \cite{efzob}. $\square$

It would be interesting to find an algebraic structure generalizing the anti-Frobenius algebras that 
corresponds to the whole set of relations (\ref{r1}), (\ref{r2})-(\ref{r5}). 

We may interpret the four index tensors $r$ and $a$ as:
 
 1) operators on $V\otimes V$, where $V$ is an $m$-dimensional vector space; 
 
 2)  elements of $Mat_m(\C)\otimes Mat_m(\C)$;
 
  3) operators on $Mat_m(\C)$.

For the first interpretation let $V$ be a linear space with a basis $e_\alpha,~\alpha=1,...,m$. Define linear
operators $r,~a$ on the space $V\otimes V$ by 
$$r(e_\alpha\otimes
e_\beta)=r^{\sigma \epsilon}_{\alpha \beta}e_\sigma\otimes e_\epsilon,\qquad a(e_\alpha \otimes e_\beta)=a^{\sigma \epsilon}_{\alpha \beta}e_\sigma\otimes e_\epsilon.
$$ 
Then the identities (\ref{r1}), (\ref{r2})-(\ref{r5}) 
can be written as 
\begin{equation}
\begin{array}{c} \label{rraa}
r^{12}=-r^{21},~~~r^{23}r^{12}+r^{31}r^{23}+r^{12}r^{31}=0,\\[5mm]
a^{12}a^{31}=a^{31}a^{12},\\[5mm]
\sigma^{23}a^{13}a^{12}=a^{12}r^{23}-r^{23}a^{12},\\[5mm]
a^{32}a^{12}=r^{13}a^{12}-a^{32}r^{13}.
\end{array}
\end{equation}
Here all operators act in $V\otimes V\otimes V$,  $\sigma^{ij}$ 
means the transposition of $i$-th and $j$-th components of the tensor
product, and $a^{ij},~r^{ij}$ mean operators $a,~r$ acting in the
product of the $i$-th and $j$-th components. 

Note that first two relarions mean that the tensor $r$ should be skew-symmetric solution of the classical associative Yang-Baxter equation \cite{Agui}.

In the second interpretation we consider the following elements from  $Mat_m(\C)\otimes Mat_m(\C)$:
$r=r^{km}_{ij} e^{i}_{k} \otimes e^{j}_{m}, \quad a=a^{km}_{ij} e^{i}_{k} \otimes e^{j}_{m},$ where $e^{i}_{j}$ are the matrix unities:
$e^j_i e^m_k=\delta^j_k e^m_i.$  Then (\ref{r1}), (\ref{r2})-(\ref{r5}) are equivalent to (\ref{rraa}), where 
tensors belong to $Mat_m(\C)\otimes Mat_m(\C)\otimes Mat_m(\C).$ Namely,   $r^{12}=r^{mk}_{ij} e^{i}_{k} \otimes e^{j}_{m}\otimes 1$ and so on. The element $\sigma$ is 
given by $\sigma =e^j_i\otimes e^i_j $.

For the third interpretation, we shall define operators $r, a, \bar r, a^{*}: Mat_{N}\to Mat_{N}$ by 
 $\quad r(x)^p_q=r^{m p}_{n q} x^{n}_{m}$,  $\quad a(x)^p_q=a^{m p}_{n q} x^{n}_{m},$ $\bar r(x)^p_q=r^{p m}_{n q} x^{n}_{m}, \quad   a^{*}(x)^p_q=a^{p m}_{q n} x^{n}_{m}.$ 
 
 Then (\ref{r1}), (\ref{r2})-(\ref{r5}) provide the following operator identities:  
$$ 
\begin{array}{c}
 r(x)=-r^*(x), \qquad r(x) r(y)=r(x r(y))+r(x) y),\\[2mm]
 \bar r(x)=-\bar r^*(x), \qquad \bar r(x) \bar r(y)=\bar r(x \bar r(y))+\bar r(x) y),\\[2mm]
  a(x) a^{*}(y)= a^{*}(y) a(x), \\[2mm]
 a^*(y a(x))=r(x   a^*(y))-r(x)  a^*(y),\\[2mm]
 a(x)a(y)=-a(r(y) x)-a(y r(x)),\\[2mm]
 a^*(a(x) y)=r( a^*(y) x)-   a^*(y) r(x),\\[2mm]
 a(y a^*(x))=-\bar r(x   a(y))+\bar r(x)  a(y),\\[2mm]
 a^*(x)a^*(y)=a^*(\bar r(y) x)+a^*(y \bar r(x)),\\[2mm]
a(a^*(x) y)=-\bar r(a(y) x)+ a(y) \bar r(x)
\end{array}
$$ for any $x,y$.  First two of these identities mean that  operators  $r$ and $\bar r$ satisfies the Rota-Baxter equation \cite{Rota} and this  fact implies also
that the new matrix multiplications $\circ_r$ and $\circ_{\bar r}$ defined by
$$
x\circ_{r}y= r(x)y + xr(y),\quad x\circ_{\bar r}y= {\bar r}(x)y + x{\bar r}(y)
$$
are associative.

 \subsection{Examples and classification of low dimensional quadratic double Poisson brackets}

It is easy to see that for $m=1$ non-zero quadratic double Poisson brackets does not exist. In the simplest non-trivial case $m=2$ the system of algebraic 
equations (\ref{r1}), (\ref{r2})-(\ref{r5}) can be solved straightforwarly. 

 {\bf Theorem 1.} Let $m=2.$ Then the following Cases 1-7 form a complete
list of quadratic double Poisson brackets up to  equivalence (\ref{basis}). 
We present non-zero
components of the tensors $r$ and $a$ only.

{\bf Case 1.}  $r^{21}_{22}=-r^{12}_{22}=1$.
The corresponding (non-zero) double brackets read 
 $$\ldb� v,v\rdb = v\otimes u - u\otimes v;$$

{\bf Case 2.} $r^{21}_{22}=-r^{12}_{22}=1,$   $a^{11}_{21}=a^{12}_{22}=1$.
The corresponding (non-zero) double brackets:
$$\ldb� v,v\rdb = v\otimes u - u\otimes v + vu\otimes 1 - 1\otimes vu, \, \ldb v,u\rdb = u^2 \otimes 1,\,
   \ldb u,v\rdb = - 1\otimes u^2 ;$$

{\bf Case 3.}  $r^{21}_{22}=-r^{12}_{22}=1,$   $a^{11}_{12}=a^{21}_{22}=1$.
The corresponding (non-zero) double brackets:
 $$\ldb� v,v\rdb = v\otimes u - u\otimes v + uv\otimes 1 - 1\otimes uv, \, \ldb u,v\rdb = u^2 \otimes 1,\,
   \ldb v,u\rdb = - 1\otimes u^2 ;$$

{\bf Case 4.} $r^{22}_{21}=-r^{22}_{12}=1$.  The corresponding (non-zero) double brackets:
 $$\ldb� v,u\rdb = v\otimes v, \,  \ldb� u,v\rdb = - v\otimes v;$$

{\bf Case 5.} $r^{22}_{21}=-r^{22}_{12}=1,$;   $a^{21}_{11}=a^{22}_{12}=1$.
The corresponding (non-zero) double brackets:
 $$\ldb� v,u\rdb = v\otimes v - 1\otimes v^2, \,  \ldb� u,v\rdb = - v\otimes v + v^2 \otimes 1,\, \ldb u,u\rdb = uv\otimes 1 - 1\otimes uv;$$

{\bf Case 6.} $r^{22}_{21}=-r^{22}_{12}=1,$;   $a^{12}_{11}=a^{22}_{21}=-1$. 
The corresponding (non-zero) double brackets:
 $$\ldb� v,u\rdb = v\otimes v - v^2 \otimes 1, \,  \ldb� u,v\rdb = - v\otimes v + 1 \otimes v^2,\, \ldb u,u\rdb = - vu\otimes 1 + 1\otimes vu;$$

 {\bf Case 7.}   $a^{11}_{22}=1$.  The corresponding (non-zero) double brackets:
 $$\ldb� v,v\rdb = u^2\otimes 1 - 1\otimes u^2.$$

 {\bf Proof.} Solving the system (\ref{r2}) for six components of the skew-symmetric tensor $r,$ we obtain the following two solutions (we present non-zero components of the tensor $r$ only): 
\begin{equation} \label{sol1}
 r^{21}_{22}=- r^{12}_{22}=x^2, \qquad r^{21}_{11}=- r^{12}_{11}=y^2, \qquad r^{21}_{12}=r^{21}_{21}=- r^{12}_{21}=-r^{12}_{12}=x y 
 \end{equation}
 and 
\begin{equation} \label{sol2}
 r_{21}^{22}=- r_{12}^{22}=x^2, \qquad r_{21}^{11}=- r_{12}^{11}=y^2, \qquad r_{21}^{12}=r_{21}^{21}=- r_{12}^{21}=-r_{12}^{12}=x y, 
 \end{equation}
 where $x$ and $y$ are arbitrary parameters. Under the transformation (\ref{basis}) the parameters in (\ref{sol1}) are changed as follows:
 $$x\to \frac{1}{\Delta}(x g_{22}+y g_{12}), \qquad y\to \frac{1}{\Delta}(x g_{21}+y g_{11}),$$
 where $\Delta=g_{22} g_{11}-g_{12} g_{21}.$ For solution (\ref{sol2}) we have 
$$x\to \frac{1}{\Delta}(-x g_{11}+y g_{21}), \qquad y\to \frac{1}{\Delta}(x g_{12}-y g_{22}).$$

For non-zero solution (\ref{sol1})  the remaining system (\ref{r3})-(\ref{r5}) for the tensor $a$ besides for zero solution has the following  two solutions:
\begin{equation} \label{sol1-1}
 a^{11}_{21}=a^{12}_{22}=x^2, \qquad a^{21}_{11}=a^{22}_{12}=-y^2, \qquad a_{11}^{11}=a_{12}^{12}=- a_{21}^{21}=-a_{22}^{22}=x y, 
 \end{equation}
 and 
 \begin{equation} \label{sol1-2}
 a^{11}_{12}=a^{21}_{22}=x^2, \qquad a^{12}_{11}=a^{22}_{21}=-y^2, \qquad a_{11}^{11}=a_{21}^{21}=- a_{12}^{12}=-a_{22}^{22}=x y. 
 \end{equation}

For  (\ref{sol2})  the system (\ref{r3})-(\ref{r5})   has the following  two solutions:
\begin{equation} \label{sol2-1}
 a^{21}_{11}=a^{22}_{12}=x^2, \qquad a^{11}_{21}=a^{12}_{22}=-y^2, \qquad a_{11}^{11}=a_{12}^{12}=- a_{21}^{21}=-a_{22}^{22}=x y, 
 \end{equation}
 and 
 \begin{equation} \label{sol2-2}
  a^{12}_{11}=a^{22}_{21}=-x^2, \qquad a^{11}_{12}=a^{21}_{22}=y^2, \qquad a_{12}^{12}=a_{22}^{22}=- a_{11}^{11}=-a_{21}^{21}=x y. 
 \end{equation}

In the case of zero tensor $r$ the remaining system (\ref{r3})-(\ref{r5}) has the following solution:
$$
a^{11}_{22}=x^4, \qquad  a^{11}_{12}=a^{11}_{21}=-a^{12}_{22}=-a^{21}_{22}=x^3 y,
$$
$$
 a^{11}_{11}=a^{22}_{22}=-a^{12}_{12}=-a^{12}_{21}=-a^{21}_{12}=-a^{21}_{21}=x^2 y^2,
$$
$$
a^{22}_{12}=a^{22}_{21}=-a^{12}_{11}=-a^{21}_{11}=x y^3,  \qquad   a_{11}^{22}=y^4
$$
with the transformation rule 
$$x\to \frac{1}{\Delta^2}(x g_{22}+y g_{12}), \qquad y\to \frac{1}{\Delta^2}(x g_{21}+y g_{11}).$$
Using (\ref{basis}), we  normalize the solutions obtained above  by $x=1,\, y=0$ and arrive at the statement of Theorem 1. 

{\bf Remark 1.} Cases 2 and 3 as well as Cases 5 and 6 are linked via involution (\ref{tran3}). 

{\bf Remark 2.} Case 1 is equivalent to the double bracket from Example 1 with $m=2.$

{\bf Remark 3.} It is easy to verify (see \cite{Agui}) that there exist only two non-isomorphic anti-Frobenius subalgebras in $Mat_2$. They are matrices with one zero 
column and matrices with one zero row.  Cases 1 and 4 correspond to them.

{\bf Remark 4.} Notice that the trace Poisson brackets for cases 2 and 4 are non-degenerate. Corresponding symplectic forms can be found in \cite{biel} (Example 5.7 and Lemma 7.1).  

{\bf Remark 5.} The corresponding Lie algebra structures on the trace space $A/[A,A]$ defining by \ref{trbr} are trivial (abelian) in all cases, except the cases 2, 3 and 4 :
$$ 
[\bar u,\bar v] = -\bar u^2 \quad ({\rm Case}\, 2), \quad [\bar u,\bar v] = \bar u^2\quad  ({\rm Case}\, 3),\quad [\bar u,\bar v] = -\bar v^2\quad ({\rm Case}\,  4).
$$
This cases give the isomorphic Lie algebra structures on $A/[A,A]$ with respect to the involutions
$u\to v,\quad v\to u$ and $u\to u\quad v\to -v.$

{\bf Example 2. } Consider the trace Poisson bracket (\ref{Poisson}) corresponding to {\bf Case 6}. Its Casimir functions are given by
$$
\mbox{tr}\,v^k, \qquad \mbox{tr}\,u v^k,  \qquad k=0,1,...
$$
where $u=x_1, v=x_2.$ Functions $\mbox{tr} \, u^i$ and $\mbox{tr} \,v u^i$, where $i=2,3,...$ commute each other with 
respect to this bracket.

The simplest integrable  ODE  system with matrix variables  corresponds to  the Hamiltonian $H=\frac{1}{2} \mbox{tr}\,u^2.$ 
This system has the following form 
\begin{equation} \label{int}
u_{t}=v u^2-u v u, \qquad v_{t}=-u v^2+v u v. 
\end{equation}
The matrix  $v^{-1} u$ is an integral of motion for this system. 
The corresponding reduction $u=v C,$  where $C$ is arbitrary constant matrix,  gives rise to known integrable model \cite{MikSok}
$$
v_t=v^2 C v-v C v^2.
$$
The cyclic reduction of the latter equation yields the non-abelian modified Volterra equation.

To study symplectic leaves for this bracket we show that the bracket is equivalent to  a pencil of
compatible linear Poisson brackets.

Let $$v=T \Lambda T^{-1}, \qquad u=T Y T^{-1},$$ where
$Y$ is a generic matrix, $\Lambda=diag(\lambda_1,...,\lambda_m),$ where $\lambda_i\ne \lambda_j$
and $\lambda_i\ne 0$, and $T$ is a  generic invertible matrix with $t_{1,j}=1$. If we fix values of the Casimir functions $\mbox{tr}\,v^k$ then $\lambda_i$ become constants. 

Consider $y_{i,j}$ and $t_{i,j},\, i>1$ as coordinates on the corresponding $(2 n^2-n)$-
dimensional Poisson submanifold. Then in this coordinates the restriction of the initial
quadratic Poisson bracket $\{,\}$ has the form $$\{,\}=\sum_{i=1}^{m} \lambda_i\,  \{,\}_i,$$
 where $\{,\}_i$ are some linear Poisson brackets.

 Describe the structure of the Lie algebra $\cal G$ corresponding to the pencil.
It turns out that $${\cal G}={\cal Y}\oplus {\cal T},$$ where
$[{\cal Y},{\cal Y}]\subset {\cal Y},$ $[{\cal Y},{\cal T}]\subset {\cal T},$ $[{\cal T},{\cal T}]=\{0\}.$
The subalgebra ${\cal Y}$ of dimension $n^2$ is generated by $y_{ij}$ and the ${\cal Y}$-module ${\cal T}$ of dimension $n(n-1)$ is generated by $t_{i,j},\, i>1$.

As an algebra ${\cal Y}$ can be considered as  a trivial central extension  of
the algebra ${\cal Z}$ spanned by $z_{i,j}=y_{i,j}-y_{i,i}$, where $i\ne j$ by by $y_{1,1},..., y_{n,n}$.

The radical of ${\cal Z}$ is spanned by $r_i=\sum_{j\ne i} \frac{1}{\lambda_j} z_{j,i}.$

The centralizer ${\cal S}$ of $r_1$ is isomorphic to ${\mathfrak gl}_{n-1}(\C)$ with $r_1$ being
central.
The isomorphism between ${\cal S}$ and $Mat_{n-1}(\C)$ is given by
$$
e^{i}_{j}\to \frac{1}{\lambda_{j}} (z_{j+1,1}-z_{j+1,i+1}), \qquad i,j=1,...,n-1,
$$
where $z_{k,k}=0$ for any $k$. Here $e^{i}_{j}$ are the matrix unities.

The radical of ${\cal Z}$ is the direct sum of two commutative ${\cal S}$-modules of
dimensions $n-1$ and 1. The first one is spanned by $v_i=r_i-r_1$.
The second is generated by $r_1$. The commutator relations between the modules is given by
$[r_1,v_i]=v_i$.

The module ${\cal T}$ is a direct sum of $n$-dimensional submodules ${\cal T}_i$
spanned by $t_{i,k},\, i>1.$ The commutator relations are
$$
[y_{i,j}, t_{k,l}]=\delta^i_l \lambda_i (t_{k,i}-t_{k,j}). \qquad \qquad \square
$$

A complete classification in the case $m=3$ based on a straightforward analysis of equations (\ref{r1}), (\ref{r2})-(\ref{r5}) seems to be a solvable but very tedious task. 
However, additional assumptions that are equivalent to a system of linear equations for components of tensors $r$ and $a$ simplifies the problem. For example, we can easily 
obtain several new examples of double Poisson brackets assuming that $\mbox{tr}\,x_2^k$ and $\mbox{tr}\,x_3^k$, where $k=1,2,...$ are Cazimir functions. One of such brackets is given by  
$$
r^{22}_{21}=r^{23}_{31}=r^{32}_{31}=-r^{22}_{12}=-r^{32}_{13}=-r^{23}_{13}=1, \qquad a^{12}_{11}=a^{22}_{21}=a^{23}_{11}=a^{32}_{31}=-a^{23}_{13}=-1.
$$
The corresponding (non-zero) double Poisson brackets:
 $$\ldb y,x\rdb = y\otimes y - y^2 \otimes 1, \qquad \ldb x,y\rdb = -y\otimes y + 1\otimes y^2,$$
 $$ \ldb z,x\rdb = y\otimes z + z\otimes y - zy\otimes 1 - 1\otimes yz,\quad \ldb x,z\rdb = - y\otimes z  -  z\otimes y + yz\otimes 1 + 1\otimes zy ;$$
 $$\ldb x,x\rdb = -yx\otimes 1 + 1\otimes yx - zy\otimes 1 + 1\otimes zy.$$

Taking $H=\frac{1}{2} \mbox{tr}\,x_1^2$ as a Hamiltonian for the corresponding trace Poisson bracket (\ref{Poisson}), we arrive at an integrable system  
$$
u_{t}=v u^2-u v u+w v u-u w v, \qquad v_{t}=-u v^2+v u v, \qquad w_t=[w,[u,v]],
$$
where $u=x_1,\, v=x_2,\, w=x_3.$
After the reduction $w=0$ this system coincides with (\ref{int}). 

Another way to construct new examples is to consider brackets homogenious with respect to any rescaling $x_i\to \mu_i x_i, \quad \mu_i \in \C$. Notice that all canonical 
forms {\bf Case 1}-{\bf Case 7} in Theorem 1 are homogeneous. When $m=3$ one of the simplest homogeneous brackets is given by 
$$
r^{31}_{22}= -r^{13}_{22}=\alpha, \qquad     a^{13}_{22}=\beta, \qquad    a^{31}_{22}=\gamma. 
$$
for some constant $\alpha,\beta,\gamma.$
The corresponding family of (non-zero) double Poisson brackets reads as:
$$\ldb y,y\rdb = \alpha (z\otimes x - x\otimes z) +\beta (xz\otimes 1 - 1\otimes xz )+\gamma (zx\otimes 1 - 1\otimes zx).$$

 \section{Compatible linear and quadratic double Poisson bracket}
 
The bi-Hamiltonian approach to integrability has been developed by F.Magri and his group \cite{magri}. It is based on the notion of compatible Poisson brackets. 
By analogy we define compatible double Poisson brackets as follows.

{\bf Definition.} Double Poisson brackets $\ldb u, v\rdb_1$ and  
$\ldb u, v\rdb_2$ on an associative $\C-$ algebra $A$ are called compatible if   
$$
\ldb u, v\rdb_1+\lambda \ldb u, v\rdb_2
$$
is a double Poisson bracket on $A$ for any $\lambda \in \C$

 The compatibility criteria for a pair of double Poisson brackets is quite similar to the
usual one:
$$
\ldb u, \ldb v,w\rdb_2\rdb_1    + \sigma\ldb v, \ldb w,u\rdb_2\rdb_1  + \sigma^2\ldb w, \ldb u,v\rdb_2\rdb_1   +
$$
$$
+ \ldb u, \ldb v,w\rdb_1\rdb_2    + \sigma\ldb v, \ldb w,u\rdb_1\rdb_2  + \sigma^2\ldb w, \ldb u,v\rdb_1\rdb_2 =0.
$$
It is clear that compatible double Poisson brackets induce (see Proposition 1) compatible trace Poisson brackets.

Consider the case when one of the brackets is a linear double bracket and another is a quadratic. 

{\bf Proposition 3.} Let $A = \C <x_1,\ldots, x_n >$. Consider the linear and the quadratic double Poisson brackets given by the  (\ref{dlin}) and (\ref{dquad}). 
Then their compatibility conditions have the following form:
\begin{equation}\label{comcond1}
b^{s}_{\alpha \gamma} a^{vu}_{s \beta}-b^{s}_{\gamma \beta} a^{vu}_{\alpha s}+
b^{u}_{s \beta} a^{vs}_{\alpha \gamma}-b^{v}_{\alpha s} a^{su}_{\gamma \beta}=0
\end{equation}
\begin{equation}\label{comcond2}
b^{s}_{\beta \alpha} r^{uv}_{s \gamma}-b^{u}_{\beta s} r^{s v}_{\alpha \gamma}-
b^{v}_{s \alpha} r^{u s}_{\beta \gamma}-b^{v}_{\gamma s} a^{u s}_{\beta \alpha}+b^{u}_{s \gamma} a^{s v}_{\beta \alpha}=0.
\end{equation}

{\bf Proof.} Straightforward verification.

Let $A$ be an $m$-dimensional associative algebra with  the multiplication law
$e_i e_j =  b_{ij}^k e_k.$
Define linear
operators $r,~a$ on the space $A\otimes A$ by $$r(e_\alpha\otimes
e_\beta)=r^{\sigma \epsilon}_{\alpha \beta}e_\sigma\otimes e_\epsilon,\qquad a(e_\alpha \otimes e_\beta)=a^{\sigma \epsilon}_{\alpha \beta}e_\sigma\otimes
e_\epsilon.$$ 
In terms of these operators acting on $A$ the compatibility conditions (\ref{comcond1}), (\ref{comcond2})  can be rewritten as 
\begin{equation}\label{comcond11}
a(x z\otimes y)-a(x\otimes z y)+a(x\otimes z)(1\otimes y)-(x\otimes 1)a(z\otimes y)=0,
\end{equation}
and
\begin{equation}\label{comcond22}
r(y x\otimes z)-(y\otimes 1) r(x\otimes z)-r(y\otimes z)(1\otimes x)-(1\otimes z)a(y\otimes x)+a(y\otimes x)(z\otimes 1) =0.
\end{equation}
The relation (\ref{comcond11}) is nothing but the cocycle condition for the Hochschild cochains $C^2(A,A\otimes A).$ Here we consider the outer bimodule structure in $A\otimes A.$

Consider the class of associative algebras $A$ such that the first and second Hochschild cohomologies with coefficients in the outer bimodule $A\otimes A$ are trivial. In particular, 
semi-simple associative algebras belong to this class. 
If $H^2(A,A\otimes A)=0$, then
\begin{equation}\label{aaa}
a(x\otimes y)=\phi(x y)-(x\otimes 1) \phi(y)-\phi(x)(1\otimes y)
\end{equation}
for some $\phi: A\to A\otimes A.$  The operator $\phi$ is defined up to the double derivations 
$$d_s : x\to (x\otimes 1)\, s-s\, (1\otimes x),$$
where $s \in A\otimes A$ is an arbitrary element.

{\bf Proposition 4.} Suppose that the tensor $a$ is defined by (\ref{aaa}).  If $H^1(A,A\otimes A)=0$, then 
any solution of (\ref{comcond22}) has the form 
\begin{equation}\label{tensorr}
r(x\otimes y)=(x \otimes 1)\psi(y)-\psi(y) (1 \otimes x) +(1\otimes y)\phi(x)-\phi(x)(y\otimes 1)
\end{equation}
for some $\psi: A\to A\otimes A.$

{\bf Proof.} Let
$$
r(x\otimes y)=\tilde r(x\otimes y)+(1\otimes y)\phi(x)-\phi(x)(y\otimes 1).
$$
It follows directly from  (\ref{comcond22})  and (\ref{aaa}) that 
$$
\tilde r(y x\otimes z)-(y\otimes 1) \tilde r(x\otimes z)-\tilde r(y\otimes z)(1\otimes x)=0
$$
If $H^1(A,A\otimes A)=0$, then
$$
\tilde r(x\otimes y)=(x \otimes 1)\psi(y)-\psi(y) (1 \otimes x) 
$$
for some $\psi: A\to A\otimes A.$ 

We denote as usual by $\sigma$ the flip  $\sigma(x\otimes y)=y \otimes x.$
It follows from $r(x\otimes y)=-\sigma \circ r(y\otimes x)$ that 
\begin{equation}\label{mubalance}
(1\otimes y)\mu(x)-\mu(x)(y\otimes 1)+(x \otimes 1)(\sigma \circ \mu(y))-(\sigma \circ\mu(y)) (1 \otimes x)=0,
\end{equation}
where $\mu(x)=\phi(x)+ \sigma \circ \psi(x)$. 

We are searching all candidates for $\mu :A\to A\otimes A$ to be a solution of \ref{mubalance} for any $x,y\in A.$
The trivial solution $\mu =0$ and hence, $\psi = -\sigma \phi$ implies the solution for $r(x\otimes y)$ in the form
\begin{equation}\label{rrr}
r(x\otimes y)=(\sigma\circ \phi(y)) (1 \otimes x)-(x \otimes 1) (\sigma\circ \phi(y))+(1\otimes y)\phi(x)-\phi(x)(y\otimes 1).
\end{equation}

If we take $\mu$ in the form $$\mu(x)=(x\otimes 1) s - s (1\otimes x), $$
where $s\in A\otimes A$ is an arbitrary skew-symmetric element : $\sigma(s)=-s.$ Then  we can straightforwardly  verify that $\mu(x)$ is a solution of  \ref{mubalance}.

In this case 
$$\phi + \sigma \psi = (x\otimes 1)s - s(1\otimes x)$$
and we can choose $\tilde\phi = \phi + (x\otimes 1)s - s(1\otimes x)$ such that 
$\psi = -\sigma\tilde\phi$ and the answer for $r(x\otimes y)$ is again given by \ref{rrr}.

{\bf Conjecture.} If $A$ is a finite unital associative algebra such that $H^1(A,A\otimes A)=H^2(A,A\otimes A) =0,$ then all solutions of \ref{mubalance} have 
the form $(x\otimes 1)s - s(1\otimes x)$ for some $s\in \Lambda^2(A)$.

We have checked the conjecture in the case of matrix associative algebra.

%Let $a_i, 0\leq i\leq n-1$ be a base of $A$ and $\mu (x) = \sum_{ij}\mu_{ij}(x)a_i\otimes a_j$ is a solution
%of \ref{mubalance} for any $x\in A.$ Then we can write $\sigma\circ\mu(u)$ in the similar way as
%$\sigma\circ\mu (y) = \sum_{ij}\mu_{ji}(y)a_i\otimes a_j.$

%Then the equation  \ref{mubalance} can be written as
%$$\sum_{ij} \lbrace a_i\otimes [\mu_{ij}(x)ya_j -\mu_{ji}(y)a_j x] -[\mu_{ij}(x)a_i y-\mu_{ji}(y)xa_i ]%\otimes a_j\rbrace =0.$$

%Taking $x=a_k, y=a_l$ on obtain
%$$\sum_m\sum_{ij}\lbrace a_i\otimes a_m [\mu_{ij}(a_k)b_{lj}^m  -\mu_{ji}(a_l)b_{jk}^m ] -[\mu_{ij}(a_k)b_{il}^m -\mu_{ji}(a_l)b_{ki}^m ]	a_m\otimes a_j\rbrace =0.$$ Here $b_{ij}^m$ is the structure contants of $A.$

%Therefore,  without loss of generality we may assume that $\mu=0$ or, the same, $$\psi(x)=-\sigma \circ \phi(x).$$

The case $a(x\otimes y)=0$ corresponds to 
\begin{equation}\label{phi}
\phi: x\to (x\otimes 1)\, s-s\, (1\otimes x),
\end{equation}
where $s \in A\otimes A$ is any fixed element. Define a tensor $r$ by formulas (\ref{rrr}) and (\ref{phi}). Explicitely, up to a constant multiplier, 
\begin{equation}\label{rur}
r(x\otimes y)=s (y\otimes x)+(x\otimes y) s-(1\otimes y) s (1\otimes x)-(x\otimes 1) s (y\otimes 1).
\end{equation}

{\bf Theorem 2.} Ler $r$ is defined by (\ref{rur}) and  $s \in A\otimes A$ satisfies the associative Yang-Baxter equation on $A$:
\begin{equation}\label{skewYB}
s^{12}=-s^{21}, \qquad s^{23}s^{12}+s^{31}s^{23}+s^{12}s^{31}=0.
\end{equation}
Then 
$$\ldb x_{\alpha}, x_{\beta}\rdb =r_{\alpha \beta}^{u v} \, x_u \otimes x_v  $$
is a quadratic double Poisson bracket on $T(A)=\C<x_1,\ldots,x_m> $ compatible with the linear bracket 
$$\ldb x_i,x_j\rdb = b_{ij}^k x_k\otimes 1 - b_{ji}^k1\otimes x_k,$$ 
where $r_{\alpha \beta}^{u v}$ are components of $\sigma r$ and  $b_{ij}^k$ are structure constants of $A.$
 
{\bf Remark 1.} We observe that in the case $a=0$ the condition \ref{comcond22} is the outer bimodule
derivation property in the {\it first argument}. That is why the quadratic double Poisson bracket from the theorem 2 can be written in following way:
\begin{equation}\label{dbtwist}
\ldb u,v\rdb = \sigma r(u\otimes v),\quad u,v\in A.
\end{equation}
Then $\sigma r$ obviously satisfies the outer bimodule derivation property in the {\it second argument}
which guaranties the Leibniz property \ref{dub3} for the double bracket defined by \ref{dbtwist}.
In other words the tensor $ R:=\sigma r : A\otimes A \to A\otimes A$ satisfies to T. Schedler conditions (\cite{Sch}):
\begin{itemize}
\item $R(u\otimes v) = -\sigma R \sigma (u\otimes v);$
\item $R^{12}R^{13} +R^{13}R^{23}- R^{23}R^{12}=0;$
\item $R$ can be considered as a derivation of  $A^{e}\otimes A^{e}-$action  on $(A\otimes A)_{l,r}$
with values in $(A\otimes A)_{i,o}$ where  $(A\otimes A)_{l,r}$ means that $A^{e}\otimes A^{e}$ acts on the
left factor of $A\otimes A$ by the first (left) $A^{e}$ and on the right factor - by the second $A^{e}:$

$$
(u\otimes u^{o})\otimes (v\otimes v^{o})(a\otimes b) = (uau^{o})\otimes (vbv^{o}).
$$
Analogously, 
$(A\otimes A)_{i,o}$ means that $A^{e}\otimes A^{e}$ acts on the
left factor of $A\otimes A$ by the inner action and on the right factor - by the outer action:
$$
(u\otimes u^{o})\otimes (v\otimes v^{o})(a\otimes b) = (vau^{o})\otimes (ubv^{o}).
$$

\end{itemize}

{\bf Remark 2.} The conditions of the Theorem 2 are satisfied for the case of finitely dimenisonal {\it quasi-triangular coboundary infinitesimal bialgebra}(\cite{Agui}). 
The conditions \ref{skewYB} mean that the algebra $A$
has also a compatible coalgebra sructure $\Delta_s :A\to A\otimes A$ such that  $\Delta_s(x) = (x\otimes 1)s -
s(1\otimes x)$ for $s\in \Lambda^2 (A).$

{\bf Remark 3.}  We observe that there is a natural class of skew-symmetric 2-tensors $s\in \Lambda^2(A)$.
Namely, M. Van den Bergh  \cite{VdB} had introduce a notion of a "momentum" map in the case of double Poisson brackets. Let us remind that there is a  distinguish 
double derivation  $\Delta : A\to A\otimes A$ such that $\Delta(a) = a\otimes 1- 1\otimes a$ for any $a\in A.$ Then the {\it moment map} for $A$ is an element 
$m\in A$ such that $\ldb m, a\rdb = \Delta(a).$  Sometimes the double derivation $H_m :=\ldb m,-\rdb$ is called a {\it Hailtonian} double vector field. The image 
of the moment map is evidently a skew-symmetric
tensor so we can take as a particular case of the previous  remark the solution
$$\mu_m(x) = (x\otimes 1)\ldb m, b\rdb - \ldb m,b\rdb (1\otimes x) = (x\otimes 1)H_m(b) - H_m(b)(1\otimes x)$$ for any $b\in A.$

{\bf Example 3.} Let $A={\rm Mat}_2(\C) = \C <x,y,z,t>.$ Then there exists a unique (up to equivalence) quadratic bracket with $a=0$ compatible with the corresponding linear one. 
This bracket has the following form:
$$
r^{12}_{23}=r^{13}_{33}=r^{14}_{43}=r^{22}_{12}=r^{22}_{24}=r^{41}_{31}=r^{42}_{32}=r^{43}_{33}=1.
$$
The remaining non-zero components of tensor $r$ are defined by the skew-symmetricity of $r: r^{ij}_{pq}=-r^{ji}_{qp}.$ 

The corresponding (non-zero) double Poisson brackets can be expressed as
$$\ldb x, y\rdb = y\otimes y; \quad \ldb x,z \rdb = -x\otimes t; \quad \ldb y, z\rdb = x\otimes y - y\otimes t;$$
$$\ldb y,t\rdb = y\otimes y; \quad \ldb z,z\rdb = x\otimes z + t\otimes z - z\otimes x - z\otimes t;$$
$$ \ldb z,t\rdb = - t\otimes x.$$

It is a straightforward verification that a  Casimir element is given by $C = x+t$ but it is impossible to restrict the brackets to the "Casimir zero level" (the traceless matrices in the representation
$A={\rm Mat}_2(\C) = \lbrace \begin{pmatrix} % or pmatrix or bmatrix or Bmatrix or ...
      x & y \\
      z & t\\
   \end{pmatrix}\rbrace :$
$$\ldb x, y\rdb = y\otimes y; \quad \ldb x,z \rdb = x\otimes x; \quad \ldb y, z\rdb = x\otimes y + y\otimes x;$$
$$\ldb y,x\rdb = y\otimes y; \quad \ldb z,z\rdb = 0;\quad \ldb z,x\rdb = x\otimes x$$
  (the "restricted" brackets are not skew-symmetric).
  
 \section{Conclusions and perspectives}
 
 We have discussed an analogue of the Lenard- Magri compatibility for linear and quadratic double Poisson brackets in free associative algebras. We have interpreted this conditions 
in terms of Hochchild cochains and 
we have proposed few examples of  solutions to these conditions. We have classified all double Poisson brackets in the case of the free associative algebra with two generators.
Our interest to the double Poisson structures was initially motivated by some examples of a non-commutative
integrability dicussed previously in \cite{MikSok} and \cite{OdRubSok1}. We are going to review a version of non-commutative Hamiltonian formalism connected the trace and double 
Poisson brackets with the
initial approach of \cite{MikSok, OlSok} in the forthcoming publications.

There are still many other interesting questions which deserve to be discussed. The natural question of a quantization the Van den Bergh construction was posed by D. Calaque (private 
comunication and see also
http://mathoverflow.net/questions/29543/what-is-a-double-star-product). Our theorem 2 gives an idea of  such a quantization for the tensor algebra associative $r-$matrix $R$ using a 
quantization (if it is known )  of the
associative skew-symmetric $r-$matrix $s$ in \ref{skewYB}. The latter can be quantized using the ideas of
\cite{GRub}. 

We have focused in this paper on the case of the free associative algebra. But the construction of double brackets was widely studied in the framework of the non-commutative 
symplectic geometry (\cite{VdB, Ginzb, CB}) aiming to describe a trace Poisson structure on quiver path algebra representations. The paper \cite{biel} proposes some $r-matrix$ 
constructions to such quadratic structures. Some of examples from \cite{biel} are coincided with our examples. We want to stress that cited paper doesn't study  general quadratic 
double Poisson brackets and the compatibility with their linear counterparts.

The original Van den Bergh construction contains also many other interesting structures and one of them is 
a {\it Quasi-Poisson} double structure ( when the double analog of the Jacobi identity \ref{dub3} "violates" 
or the "triple product" $\ldb u,v,w \rdb \in A\otimes A \otimes A$ is non-zero but are somehow controlled). See the details in \cite{VdB}. 
Recently an interesting paper \cite{MasT} had discussed the Quasi-Poisson double structures with the analogs
of  trace brackets on representations of  the group algebra $A =K(\pi)$ where the group $\pi$ is the fundamental group of a surface. The relations with the Goldman bracket, 
skein algebra and Fox multiplication were discussed. It would be interestiong to compare our tensor approach to the results of \cite{MasT}.

Finally, the last but not the least interesting subject concerns to general ( not necessary constant ) solutions of
various associative Yang-Baxter equations. The paper in progress (\cite{OdRubSok2}) contains some preliminary results in classification of parameter-dependent double Poisson 
brackets and some of new examples of such brackets.

\vskip.3cm
\noindent
{\bf Acknowledgments.}
The authors are grateful to A. Alekseev, Y. Berest, I. Burban, D. Calaque, E.B. Vinberg, M. Kontsevich,  M. Pedroni, T. Schedler,  M. Semenov-{T}ian-{S}hansky, Z. Skoda and J.- C. Thomas
for useful discussions. 
 VS and VR  are grateful to MPIM(Bonn) and AO is grateful to IHES for hospitality while the paper was written. They are grateful to MATPYL  project "Non-commutative integrable systems" 
and the ANR ``DIADEMS'' project for a financial support of VS visits in Angers. 
They were also  partially supported by the RFBR grant 11-01-00341-a. V.R. thanks to the grant FASI RF 14.740.11.0347 and the Program "COGITO"(EGIDE) of french-croatian cooperation.

When this paper was prepared to a submission we were informed that S.V. Manakov, a famous researcher who had invested a lot in the modern Integrable Systems theory, had passed away. Many of his works were a source of inspiration for us. We are dedicating this paper to his memory.


\newpage

\begin{thebibliography}{10}


\bibitem{Agui} M. Aguiar, 
\newblock{ On the associative analog of Lie bialgebras.}
 \newblock{\em J. Algebra}, {\bf 244} , 492--532, 2001.

\bibitem{biel} R. Bielawski, 
\newblock{Quivers and Poisson structures.}
 \newblock{\em arXiv:math/1108.3222}, 2011.


\bibitem{magri} F. Magri, 
\newblock{A simple model of the integrable Hamiltonian equation},
\newblock{\em J. Math. Phys.}, {\bf 19}, 1156--1162, 1978.

\bibitem{razum} Kh. S. Nirov, A.V. Razumov, 
\newblock{W-algebras for non-abelian Toda system},
\newblock{\em J. Geometry and Phys.}, {\bf 48}, 505--545, 2003.

\bibitem{Ginzb} V. Ginzburg, 
\newblock{ Lectures on Noncommutative Geometry.}
 \newblock{\em arXiv:math/0506603}, 2005.

\bibitem{GRub} D. Gurevich and V. Rubtsov,
\newblock{Yang-Baxter equation and deformation  of Associative and Lie algebras},
\newblock{\em Lect. Not. Math.}, {\bf 1510}, 47--55, 1992.


\bibitem{Rota} G.-C. Rota, 
\newblock{Baxter operators, an introduction. Gian-Carlo Rota on combinatorics},
\newblock{\em Contemp. Math.}, Birkhauser Boston, Boston MA, 1995{\bf 57}(6), 504--512.



\bibitem{MikSok}  A.V.~Mikhailov  and V.V.~Sokolov,
\newblock{ Integrable ODEs on Associative Algebras}. 
\newblock{\em Comm. Math. Phys}. {\bf 211}, 231-251, 2000.


\bibitem{OdSok1}
 A.V.~Odesskii  and V.V.~Sokolov,
\newblock{Integrable matrix equations related to pairs of compatible associative algebras},
\newblock{\em  Journal Phys. A: Math. Gen.},  2006, {\bf 39}, 12447--12456.

\bibitem{OdRubSok1}
 A.V.~Odesskii, V.N.~Rubtsov and V.V.~Sokolov,
\newblock{Bi-hamiltonian  Ordinary Differential Equations with Matrix Coefficients},
\newblock{\em  Theor. Math. Phys.},  2012, {\bf 171}, 442--447.

\bibitem{OdRubSok2}
 A.V.~Odesskii, V.N.~Rubtsov and V. V.~Sokolov,
\newblock{Parameter-dependent double Poisson brackets},
\newblock{\em  To appear.},  2012.

\bibitem{CB}
 W.~Crawley-Boevey,
\newblock{Poisson structures on moduli spaces of representations.}
\newblock{\em  J. Algebra }{\bf 325} (2011), 205�215.

\bibitem{FL}
 D.~Farkas and G.~Letzter,
\newblock{Ring theory from symplectic geometry.} 
\newblock{\em J. Pure Appl. Algebra} {\bf 125}(1998), no. 1-3, 155�190.


\bibitem{OlSok}
 P.~Olver and V.~Sokolov,
\newblock{ Integrable evolution equations on associative algebras}. 
\newblock{\em Comm. Math. Phys. }{\bf 193} (1998), no. 2, 245�268. 



\bibitem{MasT}
 G.~Massuyeau, V. ~Turaev
\newblock {Quasi-Poisson structures on representation spaces of surfaces}
\newblock{\em  arXiv:1205.4898}, 2012.

\bibitem{CProc}
 C.~Procesi,
\newblock{The invariant theory of $n\times n$ matrices},
\newblock{\em  Advances in  Math.},  1976, {\bf 19 }, 306--381.


\bibitem{VdB}
 M.~Van den Bergh,
\newblock{Double Poisson algebras },  
\newblock{\em Trans. Amer. Math. Soc.}, {\bf 360} (2008), no. 11, 5711�5769.

\bibitem{PVdW}
A.~Pichereau and G. ~Van den Weyer,
\newblock{Double Poisson cohomology of path algebras of quivers.} 
\newblock{\em J.  of Algebra} {\bf 319}, (2008), 2166�2208.

\bibitem{Sch}
 T.~Schedler,
\newblock{ Poisson algebras and Yang-Baxter equations. Advances in quantum computation}, 91�106, \newblock{\em Contemp. Math.}, {\bf 482}, Amer. Math. Soc., Providence, RI, 2009.,

\bibitem{efzob}
O.~Sokolova, A.~Zobnin.
\newblock{Anti-Frobenius associative algebras and non-abelian quadratic Poisson brackets},(Russian)
\newblock{\em Talk at the annual Lomonosov's Conference, Moscow State University, sect. Mathematics, april 2012.} to appear.

\bibitem{Rzm}
Y.~Razmyslov
 \newblock{Identities with trace in full matrix algebras over a field of characteristic zero}, (Russian), 
 \newblock {\em Izv. Akad. Nauk SSSR Ser. Mat.}, {\bf 38} (1974), 723�756. 



\end{thebibliography}
\end{document}